\documentclass{jfm}

\usepackage{graphicx}
\usepackage{epstopdf,epsfig}
\usepackage{newtxtext}
\usepackage{newtxmath}
\usepackage{natbib}
\usepackage{hyperref}
\hypersetup{
  colorlinks = true,
  urlcolor  = blue,
  citecolor = blue,
}
\usepackage{xcolor}
\usepackage{amsmath}
\usepackage{mathtools}
\usepackage{enumerate}
\usepackage[normalem]{ulem}
\usepackage{siunitx}
\usepackage{layouts}

\newcommand{\RomanNumeralCaps}[1]
\linenumbers

\shorttitle{Shape effect on ice melting in flowing water}
\shortauthor{R. Yang, C. J. Howland, H.-R. Liu, R. Verzicco, D. Lohse}

\title{Shape effect on ice melting in flowing water}

\author{Rui Yang\aff{1},
Christopher J. Howland\aff{1},
Hao-Ran Liu\aff{2},
Roberto Verzicco\aff{1,3,4},
 \and Detlef Lohse\aff{1,5} \corresp{\email{d.lohse@utwente.nl}}}

\affiliation{$^1$Physics of Fluids Group, Max Planck Center for Complex Fluid Dynamics, and J. M. Burgers Centre for Fluid Dynamics, University of Twente, P.O. Box 217, 7500AE Enschede, The Netherlands\\
$^2$Department of Modern Mechanics, University of Science and Technology of China, Hefei 230027, China
$^3$Dipartimento di Ingegneria Industriale, University of Rome `Tor Vergata', Roma 00133, Italy\\
$^4$Gran Sasso Science Institute - Viale F. Crispi, 7 67100 L'Aquila, Italy\\
$^5$Max Planck Institute for Dynamics and Self-Organization, 37077 G\"ottingen, Germany
}

\begin{document}

\maketitle

\begin{abstract}
Iceberg melting is a critical factor for climate change, contributing to rising sea levels and climate change. However, the shape of an iceberg is an often neglected aspect of its melting process. Our study investigates the influence of different ice shapes and ambient flow velocities on melt rates by conducting direct numerical simulations. Our study focuses on the ellipsoidal shape, with the aspect ratio as the control parameter. It plays a crucial role in the melting process, resulting in significant variations in the melt rate between different shapes. Without flow, the optimal shape for a minimal melt rate is the disk (2D) or sphere (3D), due to the minimal surface area. However, as the ambient flow velocity increases, the optimal shape changes with the aspect ratio. We find that ice with an elliptical shape (when the long axis is aligned with the flow direction) can melt up to 10\% slower than a circular shape when exposed to flowing water. We provide a quantitative theoretical explanation for this optimal shape, based on the competition between surface area effects and convective heat transfer effects. Our findings provide insight into the interplay between phase transitions and ambient flows, contributing to our understanding of the iceberg melting process and highlighting the need to consider the aspect ratio effect in estimates of iceberg melt rates.
\end{abstract}


\section{Introduction}
With global warming, more icebergs are breaking off from Antarctica, accelerating sea level rise \citep{scambos2017much}. Additionally, melting icebergs and sea ice floes also contribute significantly to climate and environment change, including freshwater supply \citep{huppert1978melting}, biological productivity \citep{wadham2019ice}, and carbon sequestration \citep{duprat2016enhanced}, making the understanding of iceberg melting rates crucial for comprehending the interplay between icebergs and the climate \citep{cenedese2022icebergs}. To describe the melt rate of icebergs, different parameterization models are proposed, such as: employing empirical relations for turbulent heat transfer over a flat plate to predict iceberg melt rates \citep{weeks1973icebergs}, considering meltwater-plume effect on the melt rate with ambient flows \citep{fitzmaurice2017nonlinear}, and coupling effects of turbulence, buoyant convection, and waves \citep{martin2010parameterizing}. 

Various lab experiments and direct numerical simulations have been conducted to investigate the interaction between ice melting and ambient flow in the laboratory scale, including ice melting in still and flowing water \citep{dumore1953heat,vanier1970free,hao2002heat,hester2021aspect,weady2022anomalous}, in Rayleigh-B{\'e}nard convection \citep{davis1984pattern,dietsche1985influence,favier2019,yang2023morphology}, which involves a plate being heated from below and a plate being cooled from above \citep{Lohse2010,ecke2023turbulent}, and in vertical convection \citep{wang2021ice,yang2022abrupt}. The density anomaly effect in freshwater causes distinct flow regimes and ice melting morphology under different temperatures \citep{wang2021growth,wang2021equilibrium,wang2021ice}. Salinity results in even more complex mushy layer structures \citep{worster1997convection}, where convection also occurs \citep{du2023sea}. These studies highlight the complex interplay between ice melting and ambient flow, leading to distinct ice morphologies and melt rates.

Icebergs and ice floes exhibit considerable variations in shape and size \citep{gherardi2015characterizing}, with horizontal extents ranging from a few meters to several hundred kilometers. Despite extensive research on iceberg melt rates using models, experiments, and simulations, the potential effects of the aspect ratio on the melt rate did not get much attention. One of the experimental results is that the overall melt rate strongly depends on the aspect ratio \citep{hester2021aspect,cenedese2022icebergs}. However, this effect of the aspect ratio is still poorly understood. Therefore, to enhance the accuracy of iceberg melting predictions, it is imperative to consider the aspect ratio in models of iceberg melt rates. 

In this study, we investigate the influence of the ice shape (ellipse aspect ratio) on melt rates through numerical simulations and theoretical analysis. We focus on the scenario of ice melting in cross-flow, as detailed in Methods, and neglect buoyancy effects and basal melting due to our focus on the top view of the melting process, distinguishing our work from previous studies \citep{couston2021topography,hester2021aspect}. Our aim is to understand how the aspect ratio affects ice melt rates by conducting a series of numerical simulations.  Our findings reveal that the aspect ratio plays a crucial role in the melting process, leading to substantial variations in melt rates. We further propose a theoretical model for the dependence of melt rates on the control parameters that agrees well with the simulation results. In this model, two contributions to the total melt rate are distinguished - surface contact-induced melt and advective flow-induced melt. The observed strong aspect ratio dependence can be quantitatively understood.

\section{Numerical method and set-up}
We numerically integrate the velocity field $\boldsymbol{u}$ and the temperature field $\theta$ according to the Navier–Stokes equations, neglecting the effect of buoyancy induced by temperature differences. The melting process is modeled by the phase-field method, which has been widely used in previous studies \citep{favier2019,hester2021aspect,couston2021topography,yang2022abrupt,yang2023morphology}. In this technique, the phase field variable $\phi$ is integrated in time and space and smoothly transitions from a value of $1$ in the solid to a value of $0$ in the liquid. The equations are non-dimensionalized by inflow speed $U_0$ as the velocity scale, ice effective diameter $D$ as the length scale, and the temperature difference between the ambient flow and the ice $T_0$ as the temperature scale. The non-dimensionalized quantities include three velocity components $u_i$ with $i=1,2,3$, the pressure $p$, the temperature $\theta$, and the phase field scalar $\phi$. The dimensionless governing equations read
\begin{gather*}
\nabla \cdot \boldsymbol{u}=0, \\
\partial_t \boldsymbol{u}+\nabla \cdot(\boldsymbol{u} \boldsymbol{u})=-\nabla p+\frac{1}{Re}\left(\nabla^2 \boldsymbol{u}-\frac{\phi \boldsymbol{u}}{\eta}\right), \\
\partial_t \theta+\nabla \cdot(\boldsymbol{u} \theta)=\frac{1}{{RePr}} \nabla^2 \theta+S t \partial_t \phi, \\
\frac{\partial \phi}{\partial t}=\frac{6}{5 S t C {{RePr}}}\left[\nabla^2 \phi-\frac{1}{\beta^2} \phi(1-\phi)(1-2 \phi+C \theta)\right],
\end{gather*}
where $\beta$ is the diffusive interface thickness, which is typically set to be the mean grid spacing. $C$ is the phase mobility parameters related to the Gibbs–Thompson relation. We choose $C = 1$, which may overestimate the Gibbs-Thomson effect for high curvature. Therefore we avoid too extreme values of $\gamma$, where the high curvature regions might be inaccurate. More details can be found in previous studies \citep{yang2023morphology,hester2020improved}. Simulations are performed using the second-order staggered finite difference code AFiD, which has been extensively validated and used to study a wide range of turbulent flow problems \citep{ostilla2015multiple,yang2016pnas,liu2022heat}, including phase-change problems \citep{yang2022abrupt,yang2023morphology}. The phase field method is applied to model the phase-change process, which has been widely used in previous studies \citep{favier2019,hester2021aspect,couston2021topography,ravichandran2021melting,yang2023morphology}.  

\begin{figure}
\centering
\captionsetup{width=1\linewidth,justification=justified}
\includegraphics[width=1.0\linewidth]{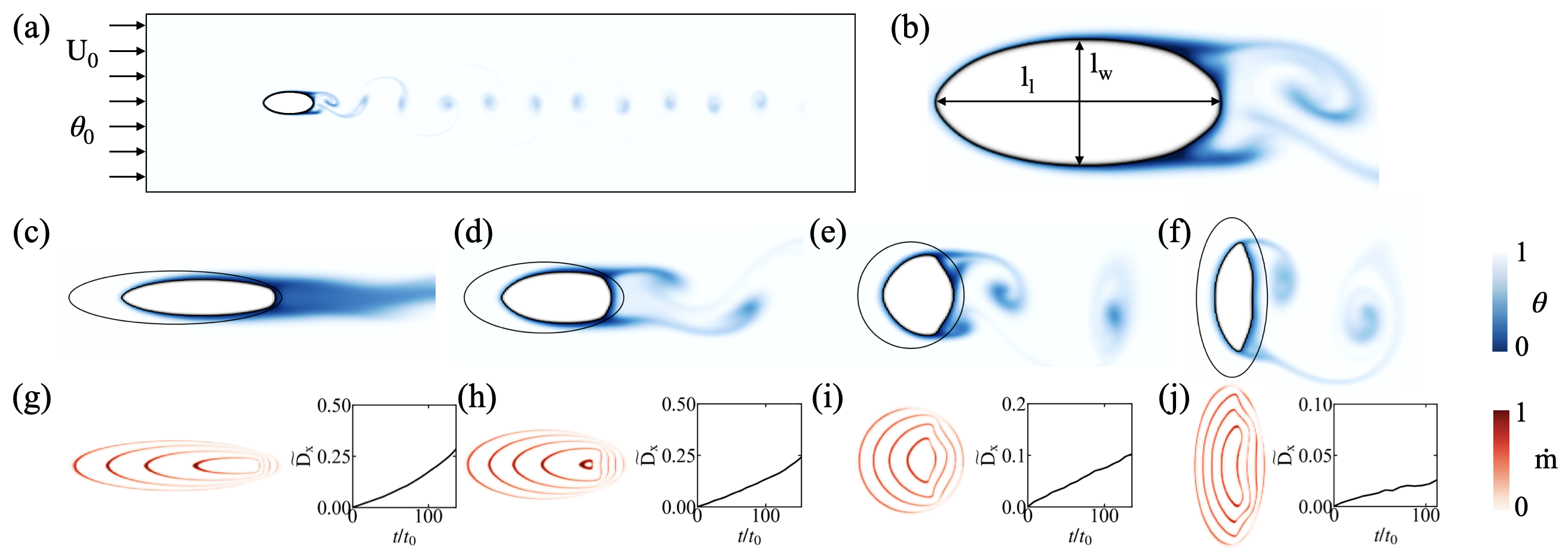}
\caption{(a). An illustration of the setup for ice melting in flowing water. The inflow is set at the left boundary with unidirectional velocity and uniform temperature $\theta_0$. (b). Zoom in on the ice object. $l_w$ and $l_l$ represent the width and the length of the object, respectively. (c)-(f) represent the snapshots of the temperature field of ice melting in flowing water with different aspect ratios $\gamma$, namely $\gamma=0.25$ (c), $0.44$ (d), $1$ (e), and $2.25$ (f). The corresponding movies are shown as supplementary materials. (g)-(j) show the snapshots of $\partial\phi/\partial t$ (see text for more details), corresponding to the cases in (c)-(f). The color represents the local melt rate over the surface at different times. Also, the corresponding plot of the shift distance $\widetilde{D}_x$ (normalized by corresponding $l_l$) of the centroids as a function of the dimensionless time is given. The mass loss rate $\dot{m}$, normalized $\partial\phi/\partial t$, is color-coded.}
\label{fig:fig1}
\end{figure}

The control parameters of the system are the Reynolds number $Re$, which is the dimensionless flow strength, the Prandtl number $Pr$, which is the ratio between kinematic viscosity $\nu$ and thermal diffusivity $\kappa$, and the Stefan number $St$, which is the ratio between latent heat and sensible heat, and the aspect ratio of the initial ice shape $\gamma$, which is defined as the ratio of its width and length and will be the focus of this article:
\begin{align}
Re&=\frac{U_0D}{\nu},& Pr&=\frac{\nu}{\kappa},& St&=\frac{\mathcal{L}}{c_p\Delta T},& \gamma&=\frac{l_w}{l_l}.
\end{align}
Here, $U_0$ is the inflow velocity, $l_w$ and $l_l$ are the initial width and length of the ice shape (shown in fig.~\ref{fig:fig1}(b)), $\mathcal{L}$ is the latent heat, $c_p$ is the specific heat capacity, $\Delta T$ is the temperature difference between inflow and the ice. Due to the large parameter space, some of the control parameters have to be fixed in order to make the study feasible. For the time being, we fix $Pr=7$ and $St=4$ as the values for water at $\rm 20^oC$, unless specified otherwise. Later we also investigate the effect of $Pr$ and $St$ independently. Our simulations cover a parameter range of $0\le Re\le 10^3$ ($Re=10^3$ corresponds to flow speed $U=5~cm/s$ and ice diameter $D=2~cm$) and $0.1 \le \gamma \le 2.25$.

\begin{figure}
\centering
\captionsetup{width=1\linewidth,justification=justified}
\includegraphics[width=0.75\linewidth]{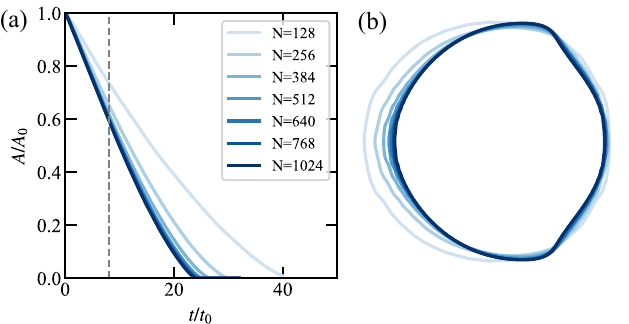}
\caption{Resolution convergence test. (a). The normalized area as a function of dimensionless time for different resolutions at $Re=400$. (b). The contour plots of the ice surface at $t=8t_0$ (gray dashed line in (a)). Based on this, our final choice for $Re=400$ is $N=512$.}
\label{fig:fig2}
\end{figure}

The flow is vertically confined by two parallel boundaries with free-slip boundary conditions for the velocity and adiabatic for the temperature (see fig.~\ref{fig:fig1}(a)). In the simulations, we prescribe an ice object with area $A_0$ and effective diameter $D=2\sqrt{A_0/\pi}$ in a domain of length $L=20D$ and width $W_z=5D$. Both two-dimensional ($L/W_z=4$) and three-dimensional simulations ($L/W_z=L/W_x=4$) are conducted. The long and short axis lengths of the ellipse are $l_l=D/\sqrt{\gamma}$ and $l_w=D\sqrt{\gamma}$, respectively. Note that the finite cross-stream dimension of the computation domain could in principle lead to blockage effects. To ensure that this does not affect our results, we performed a test for circle shapes at $Re=400$ with different $W_z$, which produced a convergent melt rate at $W_z=5D$. Initially, the simulations are run with the ice object fixed at zero velocity and $\theta_i=0$, and the melting process is turned off until the flow reaches the fully developed stage, then we turn on the melting process. The velocity and temperature at the left and right boundary are set as uniform $U_0$ and $T_0$ by a penalty force. The initial temperature of the ice is set as the melting temperature $\theta_i=0$, and the ambient fluid is set as the inflow temperature $\theta=1$. The position of the ice front is described as iso-contour $\phi=1/2$ as time evolves. A grid convergence test is shown in fig.~\ref{fig:fig2}.

\section{Shape evolution and melt rate dependence on parameters}
We first investigate the evolution of ice shapes, which depend on the initial shape and the flow rate.
Temperature fields under different initial shapes for a fixed area are shown in fig.~\ref{fig:fig1}(c-f) ($\gamma$ increases from (c) to (f)). From the temperature field, one can see the complex interaction between melting and detaching wake vortices. The ice has distinct front and back shapes - ice at the back is more deformed than at the front, due to the detaching vortices in the wake flow, which induces reduced and non-uniform heat flux. 

To study the local melt rate at the ice front, we further plot $\partial\phi/\partial t$ in fig.~\ref{fig:fig1}(g-j), whose value is zero in the solid and liquid phases. At the solid-liquid interface, its value provides insight into the local melt rate. It reveals that two parts of the ice melt faster (in reddish color): the front due to inflow, and the back due to wake flow. This melting characteristic is similar to that of dissolution \citep{ristroph2012sculpting,mac2015} and to what was formed in previous experiments on melting spheres \citep{hao2002heat}, where the front tends to melt faster than the back. This occurs because the liquid at the back of the solid is cooler than at the front due to mixing induced by the wake. We investigate the front-to-back asymmetry in the melting by plotting the distance of the ice centroid from its initial position over time in fig.~\ref{fig:fig1}(g-j). For small $\gamma$, we observe a significant shift of the centroid to the back,  while for large $\gamma$, the larger cross-section results in stronger detaching flow, which enhances melting at the back and thus prevents a significant movement of the centroid (see fig.~\ref{fig:fig1}(f)). As melting progresses, the front of the shape remains rounded, and different back shapes are obtained for different $\gamma$. These shapes can be attributed to the presence of the detaching vortices, whose shedding occurs in an oscillating manner at the top and bottom of the body. As a result, melting is favored at the top and bottom, creating a wedge-like shape.

\begin{figure}
\centering
\captionsetup{width=1\linewidth,justification=justified}
\includegraphics[width=0.8\linewidth]{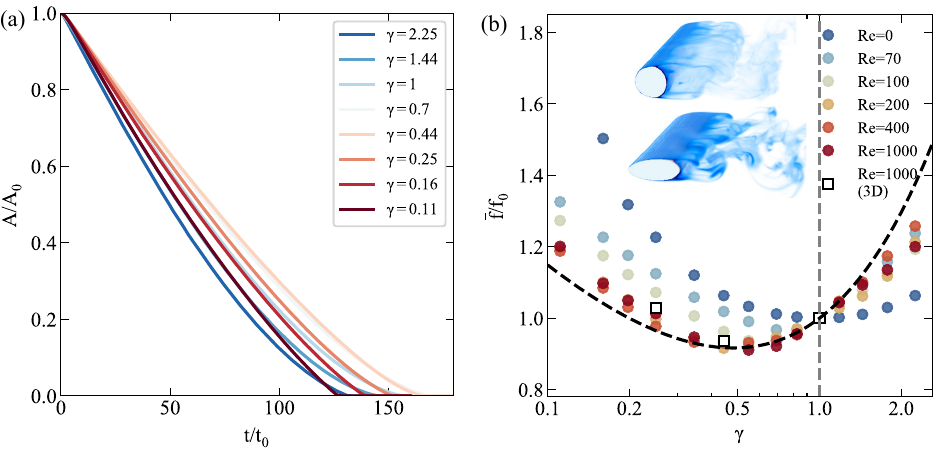}
\caption{(a). The area $A$ normalized $A_0$ as function of time normalized by $t_0=D/U_0$ for different $\gamma$ and fixed $Re=10^3$, $Pr=7$, and $St=4$. (b). Overall melt rate $\overline{f}/f_0$ (from initial to complete melt), normalized by the overall melt rate $f_0$ at $\gamma=1$, as a function of $\gamma$ for different $Re$. The inset image shows the snapshots of temperature fields for 3D simulation results ($Re=10^3$). The dashed line is the theoretical curve from eq.~(\ref{eq:tfinal}), which is expected to hold for larger $Re$.} 
\label{fig:fig3}
\end{figure}

The complicated ice-water interaction not only affects the melting shape but also the melt rate. Fig.~\ref{fig:fig3}(a) shows the normalized total area change over time for different aspect ratio $\gamma$. For increasing $\gamma$, the melt curves exhibit a non-monotonic trend, with the melt rate first decreasing and then increasing. We measure the time taken for the ice to completely melt as $t_f$, and take $f=1/t_f$ to be the mean melt rate. Fig.~\ref{fig:fig3}(b) shows the normalized melt rate $f/f_0$ as a function of $\gamma$ for different $Re$, where $f_0(Re)$ is the mean melt rate for the circles ($\gamma=1$). Here, the same trend is observed, and $\overline{f}$ depends not only on $\gamma$ but also on $Re$. In the absence of flow ($Re=0$), the minimum melt rate occurs at the optimal aspect ratio $\gamma_{min}=1$ for a circle shape. As $Re$ increases, $\gamma_{min}$ decreases until it converges to around $\gamma_{min}\approx0.5$. The same trend is also observed for 3D simulations of a cylinder with the cross-section of different aspect ratios. 

Our findings demonstrate that ice with an elliptical shape (when the long axis is aligned with the flow direction) can melt up to 10\% slower than a circle shape when exposed to flowing water. This previously neglected shape factor may have implications for accurately estimating the melt rate of icebergs in previous models that neglected it \citep{weeks1973icebergs,martin2010parameterizing,fitzmaurice2017nonlinear}. The non-intuitive nature of this phenomenon raises a further question about its physical mechanism.

\section{Theoretical model for the melt rate}
\begin{figure}
\centering
\captionsetup{width=1\linewidth,justification=justified}
\includegraphics[width=0.8\linewidth]{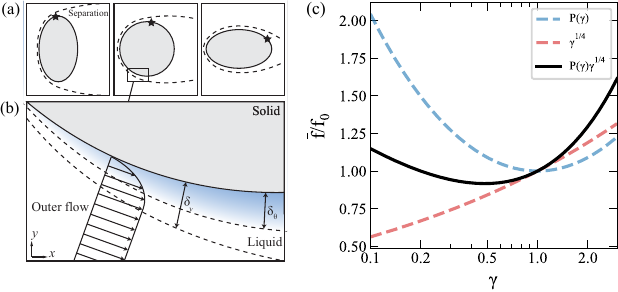}
\caption{(a). A schematic of the flow and temperature fields for different $\gamma$. The steady outer flow consists of warm water, while the attached flow near the body forms a boundary layer (dashed line) containing the melted fluid. It also shows that the separation point of flow moves downstream as $\gamma$ decreases. (b). Zoomed-in view of the velocity and temperature boundary layers, the former defined by the velocity and the latter by the temperature gradient. (c). The theoretical curve from eq.~(\ref{eq:tfinal}), including $P(\gamma)$, $\epsilon^{1/4}$, and $P(\gamma)\gamma^{1/4}$ as a function of $\gamma$. Here, Pr=7, St=4.}
\label{fig:fig4}
\end{figure}

To quantitatively explain the result, we consider the total ice mass budget for the melting process:
\begin{equation}
\frac{dA(t)}{dt}=P(\gamma,t)v_n,
\label{eq:dAdt}
\end{equation}
where $P(\gamma,t)$ is the perimeter of the ice (for 2D), and $v_n(t)$ is the surface averaged melt speed. Assuming the elliptical shape is not significantly changing with time (i.e., $\gamma(t)\approx\gamma(t=0)=\gamma$), we have the expression for the ellipse perimeter from the elliptic integral:
\begin{equation}
P(\gamma,t)=2\sqrt{\frac{A(t)}{\pi\gamma}}\int^{2\pi}_0\sqrt{1-(1-\gamma^2)\sin^2\alpha}d\alpha,
\label{eq:P}
\end{equation}
where $\alpha$ is the angle. $v_n(t)$ would be uniform everywhere in the absence of flow. The presence of a flow around the body increases the temperature gradient at the ice front which enhances the heat transfer and also makes $v_n(t)$ non-uniform as $v_n(\alpha,t)$. The flow creates a viscous boundary layer and a thermal boundary layer, respectively of thicknesses $\delta_\nu$ and $\delta_\theta$. The width ratio of the boundary layers is given by the Prandtl number; $Pr=\nu/\kappa=7$ in our case implies $\delta_v>\delta_\theta$, as illustrated in fig.~\ref{fig:fig4}(a-b). In order to understand the coupling between the shape of the object and the flow around it, we consider the Stefan boundary condition, i.e. the (dimensionless) surface-averaged melt speed $\tilde{v}_n$ is related to the surface-averaged heat flux $\overline{Nu}$:
\begin{equation}
\tilde{v}_n=\frac{v_n}{U_0}=-\frac{1}{U_0}\frac{\kappa c_p}{\mathcal{L}}\frac{\partial T}{\partial n}=\frac{\overline{Nu}}{StPrReA^{1/2}},
\label{eq:vn}
\end{equation}
where we can use the relation from the well-known scaling of thermal boundary layer thickness \citep{meksyn1961new,grossmann2004fluctuations} for the laminar boundary layer:
\begin{equation}
\overline{Nu}\sim\delta_\theta^{-1}\sim Re_{l_w}^{1/2}Pr^{1/3}/C(Pr),
\label{eq:Numean}
\end{equation}
with $Re_{l_w}=Rel_w/D$ the Reynolds number defined by the cross-section length $l_w$, $C(Pr)$ is an infinite alternating series given in \cite{meksyn1961new}. In the limit of large $Pr$, the series for $C(Pr)$ will converge to $C(Pr) = 1$. By substituting eq.~(\ref{eq:P}) and eq.~(\ref{eq:vn}) into eq.~(\ref{eq:dAdt}), we obtain an ODE which is easily solved, with the result
\begin{equation}
A(t)=A_0\left(1-\frac{t}{t_f}\right)^{\frac{4}{3}},
\label{eq:dAdtfinal}
\end{equation}
\begin{equation}
\ t_f^{-1}\sim P(\gamma)\gamma^{1/4}\frac{Pr^{-2/3}}{C(Pr)}St^{-1}.
\label{eq:tfinal}
\end{equation}
Thus the overall melt rate $\overline{f}$ depends on $\gamma$ as $\overline{f}=t_f^{-1}\sim P(\gamma)\gamma^{1/4}$, where $\gamma^{1/4}$ originates from the scaling of $\overline{Nu}$, representing the advective flow-induced melt, and $P(\gamma)$ represents the surface contact-induced melt. The contributions of $P(\gamma)$ and $\gamma^{1/4}$ are both presented in fig.~\ref{fig:fig4}(b), with $P(\gamma)$ exhibiting a symmetric curve as a function of $\gamma$ with a minimum at $\gamma=1$, which represents melting without flow. In contrast, $\gamma^{1/4}$ displays a monotonic increase in melt rate, indicating that ambient flow enhances melt rate. The overall melt rate $\overline{f}\sim P(\gamma)\gamma^{1/4}$ has a non-monotonic trend with $\gamma$, with a minimum at $\gamma_{\rm{min,theory}}=0.48$ (which is essentially the same as the value $\gamma_{\rm{min,sim}} \simeq 0.5$ observed by our numerical simulations). Fig.~\ref{fig:fig3}(b) shows the total melt rate curve, which agrees well with the numerical results obtained at the high $Re$ in our numerical results. For relatively low $Re$ and even $Re=0$, the advective flow is weak, meaning $\overline{Nu}$ is close to $1$, instead of satisfying the boundary layer relation eq.~(\ref{eq:Numean}).

In summary, the non-monotonic relation and shift of the minimum melt rate point are physically explained by the competition between the melt driven by surface contact and the melt driven by the advective flow. The former is related to the surface area, with the minimum at $\gamma=1$, while the latter is driven by the ambient flow, leading to a monotonic increase in melt rate with increasing $\gamma$. We confirmed that the assumption of a fixed shape as the ice melts is valid by comparing the measured and calculated perimeters from eq.~(\ref{eq:P}) in fig.~\ref{fig:fig5}(a). The ratio remains close to $1$, except for large $\gamma$, which is attributed to the strong wake flow (see fig.~\ref{fig:fig1}(f)). The deviation of the assumption at large $\gamma$ also explains the deviation in fig.~\ref{fig:fig3}(b) at large $\gamma$.

\section{Melt rate scalings - comparison of theory and numerical result}
\begin{figure}
\centering
\captionsetup{width=1\linewidth,justification=justified}
\includegraphics[width=1.0\linewidth]{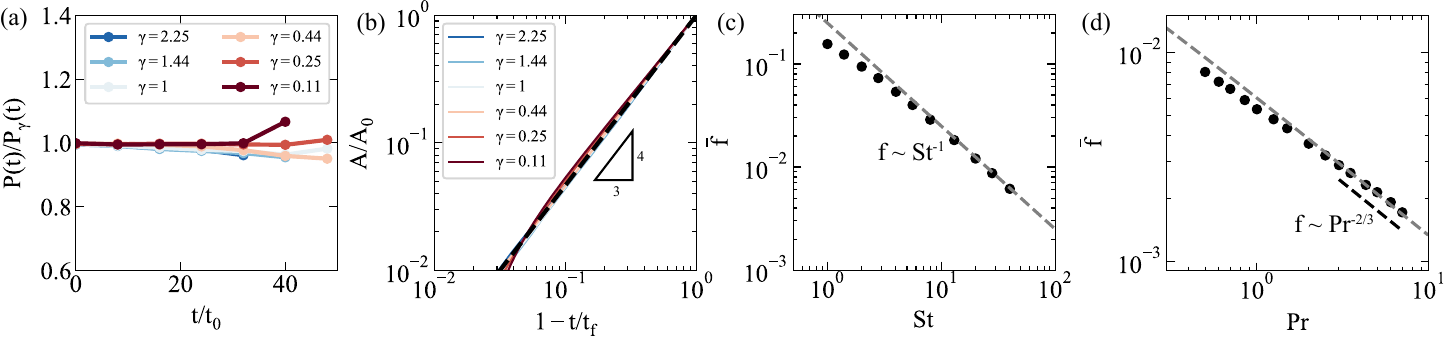}
\caption{(a). The ratio between the measured perimeter from simulations and the ideal elliptical perimeter from eq.~(\ref{eq:P}) for varying $\gamma$ at $Re=400$. The ratio close to $1$ means that our assumption is valid. (b). The normalized area $A/A_0$ as a function $1-t/t_f$ for varying $\gamma$ at $Re=400$. All curves follow the $4/3$ scaling as theoretically derived in eq.~(\ref{eq:dAdtfinal}). (c). The melt rate as a function of $St$ for fixed $\gamma=1$ and $St=4$. The dashed gray line represents $\overline{f}\sim St^{-1}$. (d). The melt rate as a function of $Pr$ for fixed $\gamma=1$ and $Pr=7$. The dashed gray line represents $\overline{f}\sim Pr^{-2/3}/C(Pr)$. }
\label{fig:fig5}
\end{figure}

In order to test the scaling law $A(t)\sim (1-t/t_f)^{4/3}$ derived in the previous section, we conducted simulations with varying $\gamma$ and fixed $Re=10^3$, $Pr=7$, and $St=4$. The evolution of the area of the object is depicted in fig.~\ref{fig:fig5}(b), along with the scaling law given by eq.~(\ref{eq:dAdtfinal}). The trends from simulations for different $\gamma$ all follow the $4/3$ scaling, in agreement with the theoretical result. Our findings demonstrate that regardless of the different physical mechanisms causing the body to ablate, whether it be through erosion \citep{ristroph2012sculpting}, dissolution \citep{mac2015}, or melting, the scaling laws associated with the vanishing of the body are analogous.

Besides the dependence of the melt rate on $\gamma$, we can also obtain the scaling dependence of the melt rate on $Pr$ and $St$ from eq.~(\ref{eq:tfinal}). To validate this scaling law, we performed simulations with varying $St$ in fig.~\ref{fig:fig5}(c) while keeping $\gamma$ fixed at $1$ and $Pr$ fixed at $7$. At large $St$ (relevant for iceberg melting in cold water), the melt rate from simulation results follows $\overline{f}\sim St^{-1}$, which is consistent with the scaling law from eq.~(\ref{eq:tfinal}). At small St, the scaling deviates from $St^{-1}$ because ice melts so fast that the melted fluid is still surrounding the ice, regardless of the ambient flow. This trend has also been observed in the study of melting in convection \citep{favier2019}. We also conducted simulations with varying $Pr$ in fig.~\ref{fig:fig5}(c) while keeping $\gamma$ fixed as $1$ and $St$ fixed as $4$. The simulation results follow $\overline{f}\sim Pr^{-2/3}/C(Pr)$ from eq.~(\ref{eq:tfinal}), implying a simplified scaling $\overline{f}\sim Pr^{-2/3}$ for large $Pr$, which agrees with our results. In brief, our simulation results for varying $St$ and $Pr$ both show good agreement with the scaling law derived in eq.~(\ref{eq:tfinal}). 

\section{Conclusions and Outlook}
Through a combination of simulations and theoretical modeling, we conducted a comprehensive investigation of the melting dynamics of a disk/elliptical shape of ice immersed in a cross-flow, taking into account the coupled dynamics between flow and ice melting for various initial shapes. Our results demonstrate that in the presence of an incoming flow, the front keeps a rounded shape during the melting process, while the back of the ice exhibits different shapes depending on the initial shape. Our simulations reveal that the presence of detaching vortices behind the ice plays a crucial role in shaping the melting pattern of the object.

Furthermore, we observed that the shape of the ice also has a significant effect on the melt rate. In the absence of flow, the circular shape ($\gamma=1$) melts more slowly than other elliptical shapes $\gamma\neq 1$. However, in the presence of an external flow, some elliptic cylindrical bodies ($\gamma<1$) melt less rapidly than circular cylinders with the same volume. This optimal aspect ratio depends on the flow strength, represented by $Re$. Our physical understanding of this dependence of the melt rate on the initial aspect ratio comes from the competition between the surface-contact-induced melt ($\sim P(\gamma)$) and the advective-flow-induced melt ($\sim \gamma^{1/4}$). By assuming a laminar boundary layer, we derived a model (eq.~(\ref{eq:dAdtfinal})and eq.~(\ref{eq:tfinal})), accurately predicting the overall melt rate and its scaling as a function of $\gamma$, $Pr$, and $St$. Our findings provide insight into the rich coupling dynamics between the ice-water interface and ambient flows and demonstrate the importance of considering ice shape in predicting melting rates.

The approach employed in this study,  which combines fully resolved direct numerical simulations and theoretical explanations,  offers the possibility to explain other phase-change problems coupled with advective flows. We caution that our findings reveal only a subset of the numerous factors that influence the ice–water system dynamics. In future investigations, it would be worthwhile to further explore additional factors such as the buoyancy force \citep{favier2019,couston2021topography,wang2021growth,yang2023morphology}, melting near sidewalls or basal walls in the presence of ambient flow, and the effect of dissolved salt \citep{huppert1978melting,yang2023salinity,du2023sea}. These topics have significant relevance to the modeling of geophysical and climatological large-scale processes. Additionally, we plan to investigate the interaction between the motion of solid objects and phase change, as this holds significant potential for further exploration since icebergs can move and rotate.

\section*{Funding}
We acknowledge PRACE for awarding us access to MareNostrum in Spain at the Barcelona Computing Center (BSC) under the project 2020235589 and project 2021250115 and the Netherlands Center for Multiscale Catalytic Energy Conversion (MCEC). We also acknowledge the support by the Priority Programme SPP 1881 Turbulent Superstructures of the Deutsche Forschungsgemeinschaf. This research was supported in part by the National Science Foundation under Grant No. NSF PHY-1748958.

\section*{Declaration of interests}
The authors report no conflict of interest.

\bibliographystyle{jfm}
\bibliography{jfmbib}

\end{document}